\documentclass[twocolumn,showpacs,preprintnumbers,amsmath,amssymb,floatfix]{revtex4}
\usepackage{graphicx}

\begin{document}


\title{Density of states and magnetoconductance of disordered Au point
contacts}

\author{L.H. Yu and D. Natelson}

\affiliation{Department of Physics and Astronomy, Rice University, 6100 Main St., Houston, TX 77005}

\date{\today}

\begin{abstract}

We report the first low temperature magnetotransport measurements on
electrochemically fabricated atomic scale gold nanojunctions.  As $T
\rightarrow 0$, the junctions exhibit nonperturbatively large zero
bias anomalies (ZBAs) in their differential conductance.  We consider
several explanations and find that the ZBAs are consistent
with a reduced local density of states (LDOS) in the disordered metal.
We suggest that this is a result of Coulomb interactions in a granular
metal with moderate intergrain coupling.  Magnetoconductance of atomic
scale junctions also differs significantly from that of less
geometrically constrained devices, and supports this explanation.

\end{abstract}

\maketitle


At low temperatures disordered metals exhibit a reduced local density
of states (LDOS) at the Fermi level, seen as a zero bias anomaly (ZBA)
in tunneling spectroscopy ({\it e.g.}  \cite{DynesetAl81PRL}).  This
is the result of disorder-enhanced electron-electron
interactions\cite{AltshuleretAl85}.  Granular metals with large
intergrain conductances ($g \equiv \langle G_{i,j} \rangle/(2e^{2}/h)
>> 1$) are predicted\cite{Efetov} to have LDOS suppressions
approaching the perturbative result\cite{AltshuleretAl85} for weakly
disordered films.  With strong disorder and geometric
constraint\cite{ButkoetAl00PRL,BielejecetAl01PRL}, the ZBA in metal
films can approach 100\%, ascribed to a correlation gap due to strong
Coulomb interactions\cite{EfrosetAl75JPC}.  Similarly, weakly coupled
granular metals ($g << 1$) should act as arrays of tunnel junctions,
with an exponentially suppressed tunneling LDOS\cite{Efetov} as $T
\rightarrow 0$.

Metallic nanojunctions (MNJs) are tools to examine geometrically
constrained, disordered metals on the nanometer scale. While clean
break junctions made in ultrahigh vacuum (UHV) have been studied
extensively\cite{Ruitenbeek}, nanojunctions between highly disordered
metals are comparatively unexplored.

We present the first low temperature measurements of atomic scale
metal junctions made by electrochemical deposition, a method proposed
for molecular electronics
investigations\cite{MorpurgoetAl99APL,LietAl00APL}.  Such metals may
be disordered by grain boundaries, ionic impurities, and surface
adsorbates.  As we reported elsewhere\cite{YuetAl03APL}, in ``large''
junctions ($G(300$~K$) >> G_{0}\equiv 2e^{2}/h$), small ZBAs are
consistent with the perturbative theory of Altshuler, Aronov and Lee
(AAL)\cite{AltshuleretAl85}.  When $G(300~K) \sim ~2e^{2}/h$, however,
low $T$ conductance suppression approaches 100\%.  Here we consider
several models and show that these junctions are atomic scale probes
of the LDOS of the disordered metal leads, which exhibit
non-perturbative, temperature-dependent LDOS corrections.  We give a
phenomenological description of the LDOS suppression, and argue that
its likely origin is the granular character of the electrodeposited
material.  Finally, these junctions exhibit nontrivial weak
localization magnetoconductances, consistent with junction size and
the hypothesis of granularity.

\begin{figure}[h!]
\begin{center}
\includegraphics[clip, width=6.5cm]{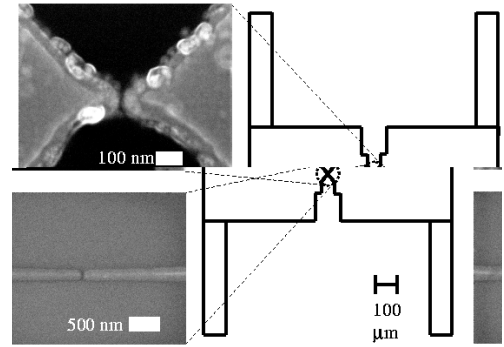}
\end{center}
\vspace{-5mm}
\caption{\small Schematic of the electrode configurations used to make
the nanoscale junctions.  At right, overall electrode geometry; at
upper left, pre-deposition image of tip geometry used for sample C;
lower left, pre-deposition image of tip geometry used for samples A
and B.  Samples A, B, and C are representative atomic scale junctions,
as discussed below.}
\label{fig:cartoon}
\vspace{-5mm}
\end{figure}

The MNJs are prepared by electrochemistry starting from electrodes
defined by e-beam lithography.  The details of sample fabrication are
available elsewhere\cite{YuetAl03APL}.  Gold electrodes 25~nm thick
with separations of $\sim$20~nm are prepared on $p+$ silicon
substrates coated with 200~nm of thermal oxide.  Electrode
configurations are shown in Fig.~\ref{fig:cartoon}.  Lumped
capacitance between each electrode and the substrate is estimated to
be $\sim$~50~pF.  The electrodes are covered by 20~nm Al$_{2}$O$_{3}$
during evaporation, limiting electrochemistry to the electrode edges.
The evaporated Au has a typical resistivity of 5~$\mu\Omega$-cm.

Additional gold is deposited using a buffered aqueous solution of
potassium cyanaurate\cite{MorpurgoetAl99APL}, while interelectrode
conductance is monitored with standard lock-in techniques.  Discrete
conductance steps on the order of $G_{0}$ are observed during junction
formation, corresponding to atomic reconfigurations.  The MNJ is grown
to a specified conductance, rinsed in deionized water, and dried with
dry nitrogen.  Through measurements on test structures, we find that
the average resistivity of the electrodeposited gold is $\sim
35$~$\mu\Omega$-cm at 4.2~K, corresponding to an elastic mean free
path of $\sim$~2.5~nm, much shorter than that of the evaporated Au.

\begin{figure}[h!]
\begin{center}
\includegraphics[clip, width=6.5cm]{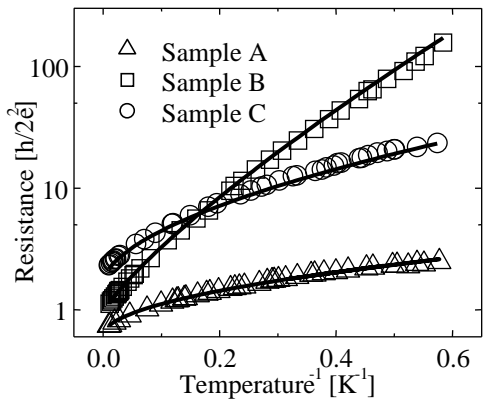}
\end{center}
\vspace{-5mm}
\caption{\small Zero bias resistance as a function of temperature for
three samples with room temperature conductances on the order of
$G_{0}$.  The form of the dependence is nonactivated.  Solid lines are
fits: (A) $0.61 \exp(1.91 T^{-0.5})$; (B) $0.91 \exp(8.06 T^{-0.8})$; (C) $1.65 \exp(3.61 T^{-0.56})$. }
\label{fig:temps}
\vspace{-5mm}
\end{figure}

Spontaneous conductance switching and its strong suppression as $T$ is
lowered suggest that the junctions consist of a small number of atoms
that can diffuse readily at room temperature.  Surviving junctions
with $G \sim 1-10 G_{0}$ are stable for tens of minutes, typical of
junctions not prepared in UHV\cite{HansenetAl00APL}.

We have successfully measured ten nanojunctions with room temperature
conductances ranging from 0.5 to 200 $G_{0}$ in a variable temperature
cryostat.  Using standard quasi-4-terminal lock-in techniques we have
measured nanojunction differential conductance $G(V,T) = dI/dV$ (and
differental resistance) as a function of temperature, dc bias voltage,
gate voltage, and magnetic field.  All nanojunctions are Ohmic up to
200~mV at 300~K.

As discussed elsewhere\cite{YuetAl03APL}, high conductance junctions
({\it e.g.} $G$(300~K)$\sim 30~G_{0}$) typically exhibit a small (15\%
at 1.8~K) ZBA, logarithmic in temperature below 30~K.  We interpret
this ZBA and its scaling with bias voltage as consistent with the AAL
perturbative DOS suppression.  An analysis of the Coulomb interaction
with a single coherent scatterer\cite{GolubevetAl01PRL} can
quantitatively explain this data\cite{Zaikin03PC}, provided the
disordered metal is granular, with an intergrain conductance on the
order of 50~$G_{0}$; such an analysis is only applicable for systems
with $G>>G_{0}$.

In atomic scale junctions prepared as above with $G(300~K) \sim 1
G_{0}$, we find ZBAs approaching 100\% conductance suppression as $T
\rightarrow 0$.  Such deep ZBAs are never seen in the larger junctions
and are present in both differential conductance and resistance
measurements.  The only difference between large and atomic scale
junctions is a brief amount of electrodeposition time; the
microstructure of the deposited material is identical.  Figure
\ref{fig:temps} shows the zero bias resistance vs. $T$ on
Arrhenius-style axes for three of these samples.

\begin{figure}[h!]
\begin{center}
\includegraphics[clip, width=6.5cm]{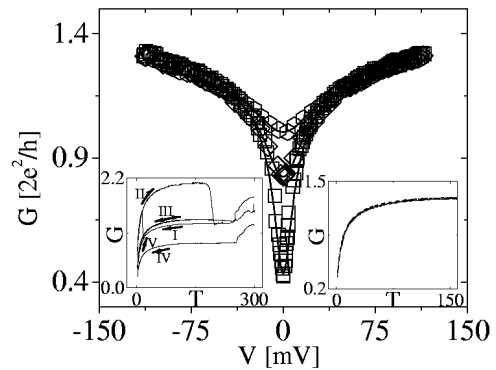}
\end{center}
\vspace{-5mm}
\caption{\small Zero bias anomaly for Sample A at (top to bottom)
20~K, 10~K, and 2~K.  Solid lines are the phenomenological model 
described in text.
Left inset:  various branches of $G(0,T)$ for this sample,
each branch corresponding to a different arrangement of 
contact atoms. Right inset: all
the branches collapse onto a single curve when each is
scaled by a multiplicative constant. }
\label{fig:zbabig}
\vspace{-5mm}
\end{figure}

We now present evidence that these large ZBAs are caused by
nonperturbative low temperature corrections to the LDOS of the
electrodeposited material.  Figure~\ref{fig:zbabig} (left inset) shows
$G(V = 0,T)$ for sample A (G(T=300K)=1$G_{0}$) as a function of
temperature cycling.  During initial cooling (trace 1), $G$ varies
little until $T< 50$~K, when the zero bias suppression begins.  At
15~K, $G$ spontaneously increases by about 0.6~$G_{0}$.  The
conductance then continues to decrease upon cooling.  When the
nanojunction is warmed (trace 2), the high temperature conductance
appears to be $\sim 2 G_{0}$.  At 220~K, however, $G$ spontaneously
decreases by about 0.85 $G_{0}$, returning near its original value.
Repeated thermal cycling and LED illumination at 2~K result in other
branches of $G$ vs. $T$ (traces 3,4,5).  {\it All} these $G(V=0,T)$
curves {\it collapse onto one curve} (right inset) when each branch of
$G(T)$ is multiplied by a non-integer constant.

The discrete changes in $G$ strongly support the idea that this
junction's low room temperature conductance ($\sim G_{0}$) is due to
the junction's atomic scale.  The addition of a single partially
transmitting channel upon cooling occurs as thermal contraction
slightly decreases the interelectrode distance.  Thermal expansion on
warming stretches the junction, and the additional channel is lost,
just as in hysteresis seen in mechanical break junction
measurements\cite{VandenbrometAl97Physica}.  These serial
rearrangements indicate that this junction was in a fortuitous regime
of stability.  In the other samples, when rearrangements took place
the junctions either broke completely or coalesced to a high
conductance state.

For each conductance branch, the ZBA is measured at several
temperatures.  Figure~\ref{fig:zbabig} shows the conductance versus
bias voltage at 2~K, 10~K, and 20~K for this nanojunction in one of
its configurations.  The same factors used to collapse the $G(V=0,T)$
branches also collapse the bias sweep data onto a {\it single set of
curves}.  It is clear that $G(V,T)$ is only {\it multiplicatively}
scaled by discrete atomic rearrangements of the small number of
conducting channels.

\begin{figure}[h!]
\begin{center}
\includegraphics[clip, width=6.5cm]{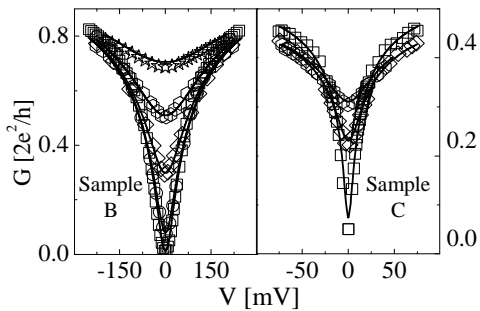}
\end{center}
\vspace{-5mm}
\caption{\small Bias sweep data for two other low conductance
junctions.  Temperatures for sample B (left) are, top to bottom, 40~K,
20~K, 10~K, 4~K, and 2~K.  Temperatures for sample C (right) are, top
to bottom, 20~K, 10~K, and 2~K.  Solid lines are calculated using the
phenomenological LDOS described in the text.}
\label{fig:zbafits}
\vspace{-5mm}
\end{figure}

We now consider possible origins of such a nonperturbative
ZBA.  A successful explanation must be consistent with:
(a) the temperature dependence of the zero bias conductance; (b) the
functional form of the ZBA vs. bias voltage and temperature; and (c)
the scaling of the ZBA data with junction conductance as shown in
Fig.~\ref{fig:zbabig}.

Coulomb blockade in an array of grains weakly coupled by tunnel
junctions can produce a ZBA\cite{PekolaetAl00APL}.  However, the
temperature dependence in Fig.~\ref{fig:temps} and the ZBA width of
our data are incompatible with such a simple model.  For a small
number of metal islands, one would also expect asymmetries or Coulomb
staircase effects, neither of which are observed.  Gate voltage sweeps
over a 100 V range produce no measurable response at any temperature
considered.  Finally, this picture would not explain the scaling in
Fig.~\ref{fig:zbabig}.

Another candidate is environmental Coulomb blockade
(ECB)\cite{DelsingetAl89PRL}.  Conduction through a single junction in
series with an environmental impedance is suppressed at low $T$ and
$V$ because tunneling electrons excite electromagnetic modes of the
environment, reducing the phase space for tunneling\cite{SCTbook}.
The $IV$ characteristics are determined by $P(E)$, the probability of
a tunneling electron to excite an environmental mode of energy $E$.
The form of $P(E)$ depends on the impedance of the environment, which
typically must be well controlled to observe ECB in single junctions.

ECB has been seen\cite{CronetAl01} in clean point contacts made using
mechanical break junctions. The environmental impedance was modeled as
the junction capacitance in parallel with the lead
resistance\cite{CronetAl01}.  Our voltage and temperature scales would
imply a charging energy $10^2 \times$ that in Ref.~\cite{CronetAl01}.
This seems unlikely given our similar geometry and greater stray
capacitance.  Furthermore, in atomic scale junctions the ECB
suppression scales with a Fano factor under changes of channel
transmittance\cite{LevyYeyatietAl01PRL}, contradicting
Fig.~\ref{fig:zbabig}.  This Fano scaling is expected to hold for a
general environmental impedance\cite{Cronthesis}, seemingly ruling out
ECB as the origin of the observed ZBAs.

A tunneling model\cite{AltshuleretAl85} is a natural approach to
analyzing the ZBA data.  One can consider a single tunnel junction
with a transmission probability $T(E)$ and an effective LDOS
$\nu(E,T)$.  The conductance $G$ normalized by the high bias / high
temperature conductance $G_{h}$ is then proportional to
$T(E)\nu(E,T)/\nu_{0}$ convolved with a thermal spreading function.
Here $\nu_{0}$ is the background effective density of states.  Our ZBA
data exhibit a considerably stronger $T$-dependence than the spreading
function; therefore a $T$-dependent LDOS is required in this model.
In this picture the transmittance is independent of $T$ and (slowly
varying in) $E$, but depends on the precise configuration of the few
atoms that make up the junction.  The LDOS is determined by disorder
``built in'' during the electrodeposition process.  This would
naturally explain the scaling in Fig.~\ref{fig:zbabig}: different
branches are equivalent to probing the same LDOS suppression using
tunnel junctions of various sizes.

The large ZBAs in the atomic scale junctions, reminiscent of
correlation gaps in highly disordered metal
films\cite{ButkoetAl00PRL,BielejecetAl01PRL}, imply a nonperturbative
LDOS suppression.  Efetov and Tscherich\cite{Efetov} consider the LDOS
in granular metals having a dimensionless intergrain conductance $g$.
They derive expressions for the LDOS in the limit $g>>1$ ($g<<1$), but
the calculated $T$-dependence is too gradual (steep) to fit our data.
This suggests that our samples fall between these extremes, into the
intermediate range of $g$ for which there is currently no analytic
expression.

A generalized treatment of corrections to electron
tunneling\cite{RollbuhleretAl01PRL} accounts nonperturbatively for
both intraelectrode Coulomb effects (the AAL LDOS correction) and
interelectrode Coulomb interactions in the presence of an
electromagnetic environment ($P(E)$ theory).  
Effective tunneling densities of states have been calculated
that agree well with experiments on spatially extended tunnel
junctions\cite{PierreetAl01PRL}.
One would expect some form of this generalized theory to apply in the
granular metal case.  We introduce an ansatz for the functional form
of $\delta \nu(\epsilon,T)$:
\begin{equation}
\delta \nu(\epsilon,T) = \nu_{0} \left(1-{\rm erf}\sqrt{\frac{-e\Gamma}{\sqrt{\epsilon^{2}+(a+bT)^{2}}}} \right)
\label{eq:ansatz}
\end{equation}
where $\Gamma$, $a$, and $b$ are sample-specific, temperature
independent parameters.  This model LDOS is able to describe
empirically the ZBA data over a broad temperature and voltage range.
The $T=0$ version of this form is derived\cite{RollbuhleretAl01PRL}
for 1d tunnel junctions, and is also equivalent to ECB in an
ultrasmall junction connected to an $RC$ transmission
line\cite{RollbuhleretAl01PRL,SCTbook}.  However, the Fano factors
discussed above for ECB in point contacts make this interpretation
difficult to reconcile with the observed scaling.  The solid lines in
Figs.~\ref{fig:zbabig} and \ref{fig:zbafits} are fits using
Eqn. (\ref{eq:ansatz}).  The relevant parameters is shown in
Table~\ref{tab:fitparam}.

\begin{table}[h!]
\begin{tabular}{|c|c|c|c|c|}
\hline
Sample & $G_{h}$ [$2e^{2}/h$] & $e\Gamma$ [J] & $a$ [J] & $b$ [J/K]\\
\hline
A & 1.51*  & $2.96 \times 10^{-22}$ & $2.33 \times 10^{-22}$ & $1.43 \times 10^{-22}$ \\
B & 1.29  & $5.11 \times 10^{-21}$ & $1.70 \times 10^{-22}$ & $6.94 \times 10^{-22}$ \\
C & 0.57**  & $6.07 \times 10^{-22}$ & $1.87 \times 10^{-23}$ & $1.57 \times 10^{-22}$ \\
\hline
\end{tabular}
\caption{\small Model parameters from Eq.~\ref{eq:ansatz}) used to reproduce the ZBA data for the samples shown in Fig.~\ref{fig:temps}. *For sample A, data from branch number 3 from Fig.~\ref{fig:zbabig} were used for the fits.  **For sample C, a slight junction rearrangement led us to use $G_{h}=0.62$ to fit data at 2~K, with the other parameters unchanged. }
\label{tab:fitparam}
\vspace{-3mm}
\end{table}

\begin{figure}[h!]
\begin{center}
\includegraphics[clip, width=6.5cm]{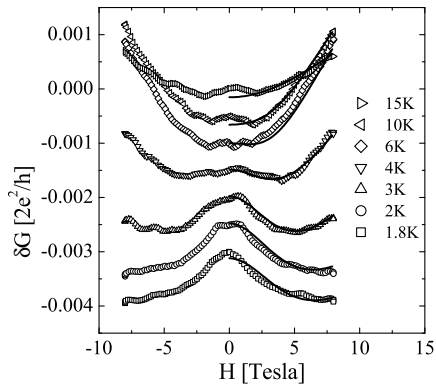}
\end{center}
\vspace{-6mm}
\caption{\small Weak localization magnetoconductance sweeps at 
several temperatures for sample B, the
junction from Fig.~\ref{fig:zbafits}.  Curves have been offset
vertically for clarity.}
\label{fig:MR}
\vspace{-3mm}
\end{figure}

We also observe magnetoconductance (MC) in atomic scale junctions that
differs from that in larger, cleaner gold systems.  Quantum MC
effects are dominated by the coherent volume centered on the
conductance-limiting junction.  Larger junctions and lead electrodes
exhibit weak antilocalization similar to that observed in clean gold
films, because the spin-orbit scattering time, $\tau_{\rm SO}$, is
more rapid than the inelastic time $\tau_{\phi}$\cite{Bergmann82ZPB}.
While $\tau_{\rm SO}$ is temperature independent, $\tau_{\phi}$
typically increases with decreasing $T$, and $\tau_{\phi} > \tau_{\rm
SO}$ for evaporated Au films below 20~K.

Figure~\ref{fig:MR} shows MC data for Sample B (data for other atomic
scale samples are similar).  Curves are offset by 0.0005$G_{0}$.
Positive MC contributions suggest that $\tau_{\phi}$ is comparable
with $\tau_{\rm SO}$, with apparently $\tau_{\rm SO} > \tau_{\phi}$
for $T> \sim$~6~K.

Quantitative analysis is difficult because the coherent volume around
the constriction is of uncertain dimensionality with respect to
diffusion and quantum coherence.  As a {\it qualitative} guide, solid
curves are fits using 1d weak localization, assuming 3d diffusion, an
effective width of 3~nm (roughly sets the field scale) and an
effective length of 5 microns (essentially a fixed numerical factor to
scale the conductance axis), and allowing $L_{\phi}$ and $L_{\rm SO}$
to vary at each temperature.  A natural explanation for a varying
$L_{\rm SO}$ would be a fixed $\tau_{\rm SO}$ but a temperature
dependent diffusion constant, $D(T)$, that decreases as $T\rightarrow
0$.  Such a dependence supports the granular metal model, where $D(T)$
would vary as intergrain interactions become important with decreasing
$T$.  We find that the ratio $(L_{\phi}/L_{\rm SO})^{2}=
\tau_{\phi}(T)/\tau_{\rm SO}$ (independent of $D$) increases with
decreasing $T$ roughly like $1/T$.  The fact that
$\tau_{\phi}/\tau_{\rm SO} \sim 1$ here suggests that $\tau_{\phi}$ is
shorter here than in larger junctions.

These first low-$T$ studies of electrochemically made atomic scale Au
nanojunctions have revealed significant departures from the properties
seen in larger and cleaner Au structures.  Atomic scale junctions
locally probe a suppressed density of states in the disordered leads.
We provide an expression that describes the data pheonomenologically,
and suggest an underlying physical origin for the effect consistent
with magnetoconductance data.  Further studies are required to
determine conclusively the physics behind the gap observed in the LDOS
and the nontrivial MC properties of this system when examined at the
atomic scale.

The authors gratefully acknowledge the support of the Robert A. Welch
Foundation and the Research Corporation.


\end{document}